\documentclass[sn-basic]{sn-jnl}

\usepackage{graphicx}%
\usepackage{multirow}%
\usepackage{amsmath,amssymb,amsfonts}%
\usepackage{amsthm}%
\usepackage{mathrsfs}%
\usepackage[title]{appendix}%
\usepackage{xcolor}%
\usepackage{textcomp}%
\usepackage{manyfoot}%
\usepackage{booktabs}%
\usepackage{algorithm}%
\usepackage{algorithmicx}%
\usepackage{algpseudocode}%
\usepackage{listings}%
\usepackage{color}
\newcommand{\ME}{M_{\rm E}}
\newcommand{\RE}{R_{\rm E}}

\newcommand{\xHe}{x_{\rm He}}
\newcommand{\Tdmx}{T_{\rm dmx}}
\newcommand{\Tone}{T_{1\,\rm bar}}

\definecolor{nnblau}{rgb}{0.0,0.0,0.8}
\definecolor{nnviol}{rgb}{0.6,0.0,0.8}

\begin{document}

\title[Article Title]{Atmospheric Helium Abundances in the Giant Planets}


\author*[1,2,3]{\fnm{Nadine} \sur{Nettelmann}}\email{nadine.nettelmann@gmx.de}

\author[3]{\fnm{Marina} \sur{Cano Amoros}}

\author[3]{\fnm{Nicola} \sur{Tosi}}

\author[1]{\fnm{Ravit} \sur{Helled}}

\author[2]{\fnm{Jonathan J.} \sur{Fortney}}

\affil*[1]{\orgdiv{Institute for Computational Science}, \orgname{University of Zurich (UZH)}, 
\orgaddress{\street{Winterthurerstr.~190}, \city{Zurich}, \postcode{8057}, \country{Switzerland}}}

\affil[2]{\orgdiv{Department of Astronomy and Astrophysics}, \orgname{University of California (UCSC)}, 
\orgaddress{\street{High St 1156}, \city{Santa Cruz}, \postcode{95064}, \state{CA}, \country{USA}}}

\affil[3]{\orgdiv{Institute of Planetary Research}, \orgname{German Aerospace Center (DLR)}, \orgaddress{\street{Rutherfordstr.~2}, \city{Berlin}, \postcode{12489}, \country{Germany}}}

\abstract{
Noble gases are accreted to the giant planets as part 
of the gas component of the planet-forming disk. While 
heavier noble gases can separate from the evolution of the 
hydrogen-rich gas, helium is thought to remain at the 
protosolar H/He ratio $Y_{\rm proto}\sim 0.27$--$0.28$. However, 
spacecraft observations revealed a depletion in helium 
in the atmospheres of Jupiter, Saturn, and Uranus. 
For the gas giants, this is commonly seen as indication 
of H/He phase separation at greater depths.

Here, we apply predictions of the H/He phase diagram and three H/He-EOS to compute the 
atmospheric helium mass abundance $ Y_{\rm atm}$ as a result of H/He phase separation.  

We obtain a strong depletion $Y_{\rm atm}<0.1$ for the ice giants if they are adiabatic. 
 
Introducing a thermal boundary layer at the Z-poor/Z-rich compositional 
transition with a temperature increase of up to a few 1000 K, we obtain a 
weak depletion in Uranus as observed. Our results suggest dissimilar internal 
structures between Uranus and Neptune. An accurate in-situ determination of 
their atmospheric He/H ratio would help to constrain their internal structures. 
This is even more true for Saturn, where we find that any considered
H/He phase diagram and H/He-EOS would be consistent with any observed value.
However, some H/He-EOS and phase diagram combinations applied to both Jupiter and Saturn 
require an outer stably-stratified layer at least in one of them.
}


\maketitle

\section{Introduction}

Many planets the size of the ice giants Uranus and Neptune have been detected. However, Uranus and Neptune
are the only known planets with deep hydrogen-helium atmospheres that are rich in volatiles and cold enough
for processes like condensation and phase separation to occur. On Uranus and Neptune we can study how such
processes influence the atmospheric composition, interior structure, and thermal evolution.

As the two remote ice giants Uranus and Neptune have the coldest tropospheres and perhaps also the 
coldest interiors of any known orbiting planet, one may consider them as end-members in the planetary landscape. 
However, they may unify a number of important planet properties and therefore be of central importance for our
understanding of the diverse population of planets.

One such important property may  be H/He phase separation. The weakly irradiated gas giant Jupiter 
is partially shaped by the phase separation of hydrogen and helium as indicated by its atmospheric 
helium depletion \citep{Niemann98,Zahn98}. If the gaseous H/He-component is present in the deep interiors 
of Uranus and Neptune, H/He phase separation may also occur there, but has not been studied in print for 
the ice giants before. 

The landscape is further populated by strongly irradiated hot Jupiters and warm Neptunes.
Their close orbits suggest that they formed farther out and migrated inward by disk-planet interaction.
For Uranus and Neptune, the Nice model of Solar System formation suggests that the ice giants formed closer 
to the Sun and migrated away later by interaction with a disk of planetesimals \citep{Tsiganis05}.
Therefore it is possible that the formation locations of warm Neptunes and Uranus and Neptune
are not that different.
From accurately measured planet masses ($M_p$) and radii ($R_p$) a bulk heavy element enrichment 
of extrasolar giant planets that increases toward lower-mass planets has been inferred \citep{Thorngren16} 
under the assumption of adiabatic interiors. With at least 65\% of their mass in heavy elements, 
Uranus and Neptune fit well in the exoplanet $M_p$--$Z_p$ relation, where $Z_p$ is a planet's heavy 
element mass fraction. Thus it could also be that planets of that mass and size share
similar bulk heavy element enrichment wherever they formed. 

Within current observational uncertainties, Uranus and Neptune are extraordinarily similar to each
other in atmospheric composition, gravity, magnetic field, rotation rate, and atmospheric temperature. 
These constraints are not available for exoplanets, however, one may assume that an exo-equivalent to 
Uranus and Neptune by mass and radius shares a similar structure. It is therefore important to identify
which structure properties they may have in common and which may be influenced by the thermal 
profile and evolution, as those will differ with the different irradiation levels they receive.

The most commonly found type of planets has even lower masses of 5--10$\:\ME$, compared to the 
$14.5\:\ME$ and $17.2\:\ME$ of Uranus and Neptune. These so-called sub-Neptunes in the 2--3 $\RE$ 
radius range occur at an average rate in the Galaxy of $\sim$0.5 planets per Sun-like star with a 
factor of $\sim$2 uncertainty \citep{Hsu19,Kunimoto20}.
Whether sub-Neptunes are mostly made of volatiles or of rocks with an extended gaseous atmosphere 
is one of the fundamental questions in planetary science and clearly not to be resolved with
further $M_p$--$R_p$ measurements alone. The same major question applies to Uranus and Neptune. 
Knowlege of bulk-$Z$, the ice:rock ratio, and the interior $Z$-distribution yields information 
on the location of formation and the conditions in the planet-forming disk. As an advantage over 
exoplanets, tracing the motion of satellites around Uranus and Neptune has allowed us to place 
constraints on their interior $Z$-distribution through the density distributions that fit the gravity 
field measurements \citep{Nettelmann13,Movshovitz22,Neuenschwander22}. 

Interior models of Uranus and Neptune constrained by the gravity data require a H/He-rich atmosphere and a
high-$Z$ deep interior, but gravitational harmonics measurements do not allow us to distinguish between a
more rocky or a more icy deep interior, nor to determine whether the transition is sharp 
or gradual \citep{Helled20}.
Depending on the temperature profile and the assumed ice:rock ratio, up to $24\%$ H-He could exist
in their deep interiors \citep{Helled11, Nettelmann13}. Thus as in Jupiter, H/He-demixing could occur there
and influence the observable atmospheric composition.

Direct measurements of element abundances are needed to place additional constraints.
However, condensible volatiles are not necessarily well mixed even deep in the troposphere. As revealed
by the Juno Orbiter, Jupiter's ammonia abundance remains variable even down to $\sim 30$ bars,
which is well below the ammonia cloud level \citep{Guillot20}.

Fortunately, noble gases like helium are thought to be well-mixed in the troposphere. Unfortunately the
mass spectrometer aboard the Cassini spacecraft stopped working before the troposphere of Saturn
was reached. So far, the only accurate noble gas abundance measurements in a giant planet's atmosphere
are those by the Galileo entry probe at Jupiter \citep{Atreya03}. This in-situ measurement has been and 
still is highly valuable for our learning process about Jupiter's interior structure and evolution 
as well as the behavior of hydrogen and helium at high pressures. The entry probe measured a higher
 He abundance than inferred from the previous Voyager occultation and remote sensing data. This makes 
the Voyager values for Saturn \citep{CG00} and the ice giants questionable.

In this work we address the helium abundance in the atmospheres of the giant planets in the Solar System.
We consider the possibility that the atmospheric He abundances in Uranus and Neptune differ from
the \textit{final} Voyager values, which are at a protosolar level for Uranus and an
enhanced level for Neptune \citep{Atreya20}.  We assume that H/He exists in their deep interiors,
where it can undergo phase separation and influence the He abundance above in the molecular-hydrogen
region \citep{SS77a,SS77b}. We thus aim to predict the atmospheric helium abundance as a
result of H/He phase separation, but we also caution that our interior model assumptions may be too
simplistic and that an inversion from a future He abundance measurement to internal structure 
will be non-unique. Yet, different structure models will reflect on a variety of observational parameters. 
Should an optimization effort be conducted about which parameters should be measured in order to rule-out 
certain models, this work aims to make the point that the atmospheric helium abundance be part of the study.

In Section \ref{sec:methods} we describe the three H/He phase diagrams that we use to calculate 
the equilibrium He abundance after phase separation (\ref{sec:HHeDiagram}), the interior model 
assumptions (\ref{sec:methProfiles}), and the three Equations of State (EOS) that we use to compute 
planetary temperature-pressure ($P$--$T$) profiles (\ref{sec:eos}). In Section \ref{sec:obsHHe} 
we review the observational He/H abundance measurements. Section \ref{sec:results} presents the results 
for the atmospheric He abundances. In Section \ref{sec:discussion} we discuss what a putative atmospheric 
probe measurement might imply for the internal structure of the outer planets. Our main points 
are summarized in Section \ref{sec:summary}. In Appendix \ref{sec:apxBLM} we describe the construction
of the BLM21-N23 phase diagram.

\section{Methods}\label{sec:methods}

\subsection{H/He phase diagram}\label{sec:HHeDiagram}

We apply three H/He phase diagrams. Among them are
the two published H/He phase diagrams that provide the demixing temperature
$T_{\rm dmx}(P;x_{\rm He})$ as a function of pressure $P$ and on a dense grid of He particle concentrations
$x_{\rm He}=N_{\rm He}/(N_{\rm He}+N_{\rm H})$ \citep{LHR09,LHR11,SR18}. Hereafter, they are referred to 
as LHR0911 and SR18, respectively. The third H/He phase diagram is derived from the experimental
data of \citep{Brygoo21}. Our construction thereof is described in Appendix 
\ref{sec:apxBLM} and it is referred to as BLM21-N23.

The helium mass fraction is obtained as $Y=4 x_{\rm He}/(1+3x_{\rm He})$. We assume that water, if present, 
is fully dissociated and contributes two H atoms per H$_2$O molecule. This upper limit to the number 
of H atoms applies if water is in the plasma phase, which occurs at typical H/He demixing pressures 
of 1--2 Mbar for temperatures $T\gtrsim 5000\,$K \citep{Redmer11}. The adiabats of Uranus and Neptune
run close to the super-ionic/plasma phase transition. If the interior is warmer than along an adiabat 
the deep adiabats could indeed pass through the plasma phase.

The LHR0911 H/He phase diagram is obtained from DFT-MD simulations of H-He systems.
Depending on the $P$--$T$ conditions chosen, phase separation into a He-rich phase and a He-poor phase can
directly be seen in the simulation box. However, this phenomenon occurs in the simulation only
if the number of particles in the box, which has periodic boundary conditions, is sufficiently high.
For example, \citet{LHR11} observe demixing in their simulation at 6000 K and 24 Mbar 
when using a high number of electrons like 2048.  \citet{Chang23} use a machine-learning approach
to directly simulate H/He demixing in even larger systems of up to 27,000 particles. Unfortunately, 
the direct simulation of demixing is computationally highly demanding. Therefore, the common method 
of choice is to simulate a mixed H-He system with much fewer particles where the then stronger
thermal fluctuations may prevent the system from adopting a more ordered demixed state. 
\citet{LHR11} and \citet{SR18} run their simulations for the complete H/He demixing diagram with 64 electrons.

Preference for mixing or demixing and the two equilibrium He-concentrations in the latter case
are then determined according to the thermodynamics criterium that a system will strive to
reach a state of minimum Gibbs free energy $G(x;P,T)=U(x)+PV(x)-TS(x)$. This free energy is related 
to the enthalpy $H$ via $G=H-TS$. Its differential reads $dG(x) = V(x)dP -S(x)dT$, or 
$dG(x) = dH(x) -S(x)dT - TdS(x)$.

In practice, the energy difference $\Delta G$ between the assumed mixed state $G(x)$ and the
superposition of the fully demixed states, $xG(0) + (1-x) G(1)$ is calculated at a given temperature $T$
and pressure $P$. If the Gibbs free energy of mixing $\Delta G(x)$ shows two minima, this is interpreted
as preference for demixing and a double-tangent construction is applied to determine the equilibrium
abundances $x_A$ on the He-poor side and $x_B$ on the He-rich side. To determine the minima and the
double-tangent accurately, a Redlich-Kister Fit is applied to $G(x; P,T)$ at every $P,T$ point 
\citep{LHR09,SR18}.

At constant temperature, $-d(TS)$ reduces to $-T\:dS$. The entropy of the mixed state is
$S(x)=xS(0) + (1-x)S(1) + S_{\rm mix}$. In the difference $\Delta S$ between the assumed mixed state $S(x)$
and the superposition of the fully demixed states, the first two terms cancel so that only
the entropy of mixing $S_{\rm mix}$ survives. $\Delta G$ at constant temperature  can thus be
calculated as $\Delta G(x) = \Delta H(x) - T\Delta S_{mix}(x)$. The enthalpy $H(x; S,P)=U(x)+PV(x)$
is obtained from the thermal EOS $P(x; V,T)$ and the internal energy $U(x; V,T)$. Both $U(T,V)$
and $P(T,V)$ are readily accessible from the DFT-MD simulations but depend on the exchange-correlational
functional chosen. The choice of the exchange-correlational functional and the approximation used
for $S_{\rm mix}$ therefore matters.

Using the ideal entropy of mixing is the simplest approach as it  neglects interparticle interactions.
A more sophisticated method has been elaborated for H-He mixtures. \citet{Morales13} perform
thermodynamic integration to relate calculated entropy to well-known cases at less extreme $P$--$T$ conditions
via the coupling-constant integration method. This approach was also adopted for the SR18 diagram.
While the difference between both approaches is negligible for He-particle concentrations up to 0.2
\citep{McMahon12}, to be compared to the protosolar value $x_{\rm He}=0.086$, the non-ideal contributions
strongly influence the Gibbs free energy of mixing at high He concentrations and thus the presence and 
location of minima \citep{McMahon12}.

H/He demixing is thought to be driven by the metallization of hydrogen. For a detailed review of 
investigation of hydrogen metallization and in the presence of helium, see \citet{McMahon12}. 
The vdw-DF1 functional yields metallization pressure at $\sim 2$ Mbars. This is in good agreement 
with the dynamic compression experiments performed at the National Ignition Facility using laser energy 
to near-isentropically compress D samples \citep{Celliers18}. In these experiments, the insulator-to-metal 
transition is taken to be  a region of steepest slope in the measured reflectivity. A value of 2 Mbar goes 
in the direction of the 3 Mbar inferred from a jump in reflectivity observed upon compression of multiply 
shocked D samples using the $Z$-machine at the Sandia National Laboratory \citep{Knudson15}.
The vdw-DF1 functional yields a slightly stiffer Hugoniot than the PBE functional in better agreement
with the early gas gun data \citep{KD17}.

A third series of experiments used laser-heated diamond-anvil-cells (DAC) to compress hydrogen. Here,
a first-order insulator-metal transition is seen at 1-1.7 Mbar in agreement with the DFT-MD simulations
when using the PBE functional \citep{Zaghoo16}. These experiments also reproduce the 
maximum compression ratio of 4.5 at 50 GPa along the principal Hugoniot as measured at the Z-machine \citep{KD17}.
However, the PBE-functional based Hugoniot maintains a rather high compressibility toward lower
pressures at 30-40 GPa where the experimental data (gas gun, Z-machine) suggest a stiffer EOS.
For reviews of recent work on hydrogen metallization and in particular under giant planet conditions 
we refer the reader to \citet{Goncharov20} and \citet{Helled20H}.
To summarize, the LHR0911 data set uses the PBE functional and ideal entropy of mixing. The \citep{Morales13} 
data set is built on the PBE functional and the entropy of mixing including interaction effects 
(non-ideal entropy of mixing). SR18 use the vdw-DF1 functional and non-ideal entropy of mixing.

\begin{center}
\begin{figure}
\rotatebox{270}{\includegraphics[width=0.68\textwidth]{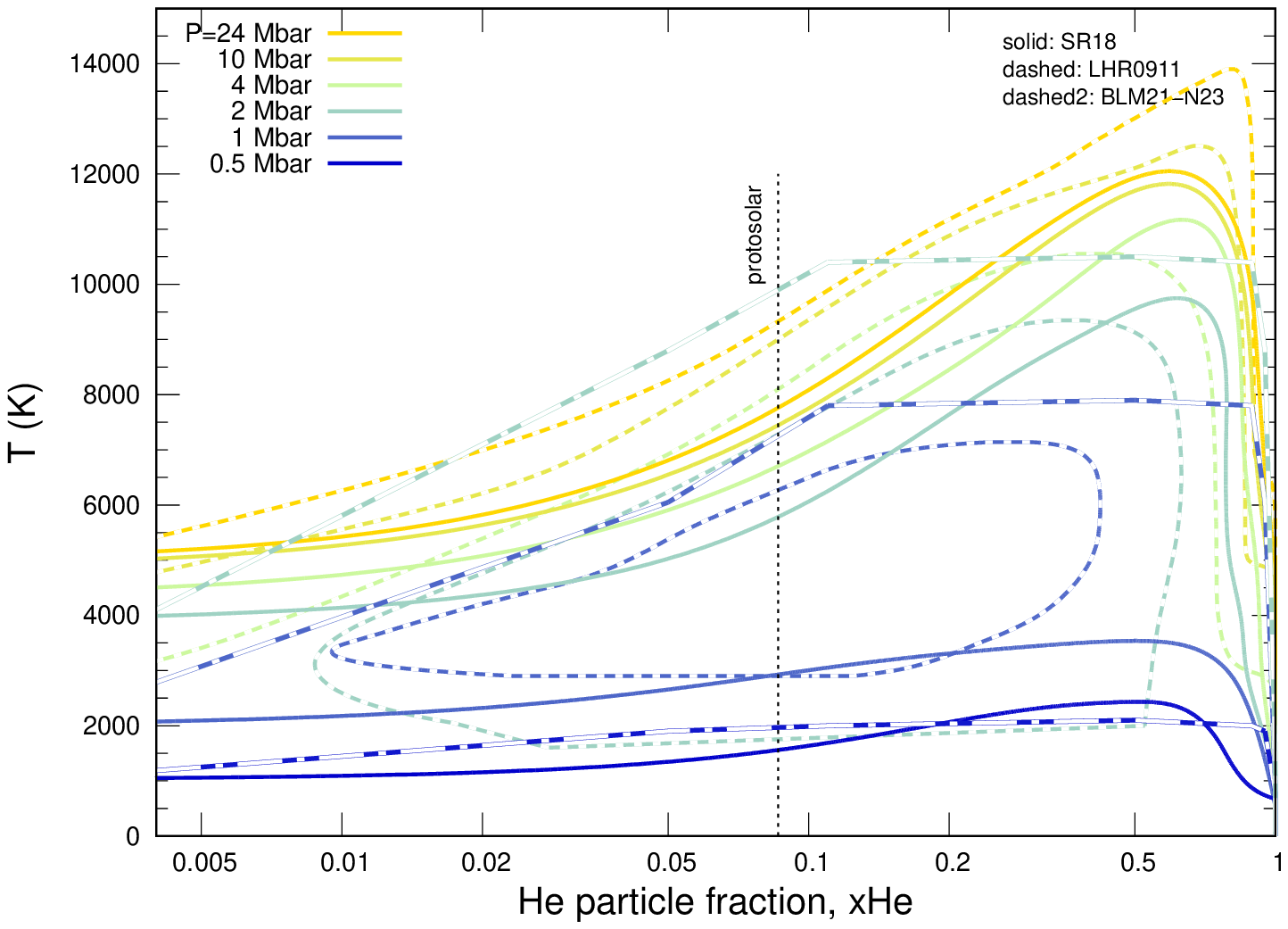}}
\caption{\label{fig:Tdmx}
The phase diagrams of LHR0911 (dashed) and SR18 (solid), and our version of BLM21 (dashed2) 
for 0.6, 1, 2 Mbars. 
The vertical dotted black line highlights the protosolar He/H ratio. Color codes isobars as labeled.}
\end{figure}
\end{center}

In Figure \ref{fig:Tdmx} we compare the three phase diagrams in terms of the
demixing temperatures $\Tdmx(P; \xHe)$. As one can see, differences in $\Tdmx$ between LHR0911 and
SR18 easily amount to 1000--1500 K, with the LHR0911 model yielding higher demixing temperatures wherever 
it predicts demixing at all.

The SR18-based $T(P)$ curve at protosolar abundance (Fig.~\ref{fig:demixad}) transitions smoothly 
into the MonteCarlo simulations results of \citet{Schouten91} and further extrapolates well 
to the DAC experimental result at a few GPa by \citet{Loubere91}.
Despite its relatively high demixing temperatures, even the LHR0911 data set does not reach a 
temperature of 10,000 K at 1.5 Mbar but rather of around 6000 K. Such a high temperature value was 
measured at the drop in the reflectivity, which was interpreted as a sign of crossing the phase boundary
in dynamically compressed H/He samples \citep{Brygoo21}. 
On the other hand, temperatures along the phase boundary for $x_{\rm He}=0.11$ (Y=0.33) used in the 
experiment fall steeply toward lower pressures. As described in Apx.~\ref{sec:apxBLM}, we also assume 
a steep decrease for lower He-concentrations. As a consequence, the demixing temperature at 1 Mbar and 
a protosolar He-concentration of our version of the BLM21 data comes close to that of the LHR0911 diagram.
Overall, Figure \ref{fig:Tdmx} indicates there may be significant uncertainties in the H/He phase diagram.

\subsubsection{Equilibrium particle concentration $x_A$}\label{sec:xA}

Suppose there is a parcel of protosolar H/He ratio $x=x_{\rm proto}$ in the interior of the planet at 2 Mbars
and 5000 K.  The parcel under these conditions would rain-out He-rich droplets until a concentration of
respectively $\xHe\sim 0.03$ (LHR0911) or $\xHe=0.06$ (SR18) is reached, see Figure \ref{fig:Tdmx}.
This is the equilibrium concentration $x_{A}$ on the phase boundary. It can be determined for any
$P,T$-point where $T(P;x)\leq T_{\rm dmx}(P;x)$.

\subsection{Planet interior profiles}\label{sec:methProfiles}

We compute interior $P$--$T$ profiles, not interior models. We assume background interior
heavy element abundances $Z$ that span a wide a range guided by previous series' of models 
for Uranus and Neptune \citep{Helled11,Nettelmann13,Nettelmann16,Scheibe21} Saturn 
\citep{Nettelmann13,Ni20,Militzer19,MaFu21}, and Jupiter \citep{Nettelmann21,Miguel22,Howard23J,Militzer23}.
For the ice giants we consider two cases, an adiabatic case and one with a thermal boundary layer (TBL). 
The latter, non-adiabatic case is described in Section \ref{sec:methTBL}. 

In the adiabatic case, we compute $P$--$T$ profiles under the assumption that the interior 
has a classical three-layer structure with an outer, heavy-element-poor envelope of water 
abundance $Z_1$, an inner, water-rich envelope of water abundance $Z_2$, and a rocky core. The 
transition between the envelopes occurs at a transition pressure $P_{1,2}$. If $P_{1,2}$ 
and the water abundances $Z_1$, $Z_2$ are adjusted to yield density distributions that 
are consistent with the low-order gravitational harmonics $J_2$, $J_4$, one can find for Uranus
$8 < P_{1,2}  < 20$ GPa, $0.01 < Z_1 < 0.2$, and $0.87 < Z_2 < 0.95$, for Neptune $5 < P_{1,2} < 150$ GPa, 
$0.01<Z_1<0.65$, and $0.7<Z_2<1$, and for Saturn $P_{1,2}\sim$0.5 to few Mbars,
$Z_1$ up to a few times solar, $Z_2$ around $Z_1$ or up to 0.5 if a deep-seated gradient is allowed for.
The situation for Jupiter is unclear because, except for few models based on MH13-H/He-EOS 
\citep{Militzer23}, HG23+MLS22 H/He-EOS \citep{Howard23J}, or CD21 but then with enhanced interior 
temperatures \citep{Debras19}, adiabatic models constrained by the tight Juno gravity data fail 
to reproduce Jupiter's observed surface temperature. Nevertheless, current Jupiter models share  
the property of a low outer envelope metallicity of 0--$2\times$ solar, $Z_2$ around $Z_1$ or slightly 
enriched up to $Z\simeq 0.2$, and $P_{1,2}$ in the 1--20 Mbar range. Whether the deep interior enrichment 
in Jupiter occurs gradually or sharply cannot be decided from the gravitational harmonics data alone. 
The Juno mission has revolutionized the label for a deep region of enhanced-$Z$.
It is now called a \emph{dilute} core \citep{Wahl17J,MaFu21}.

The way the above described models matter for this work is as follows: In Uranus and Neptune, 
the He rain region is located in the heavy element-rich deep envelope. 
We vary the water mass fraction $Z_{\rm H2O}$ between 0 (rocky case) and 0.90.
In Jupiter, He-rain occurs in the outer, heavy-element poor envelope.
In Saturn, the He-rain region reaches across the two envelopes of potentially slightly 
different but generally low $Z$. A level $Z\sim 0.1$ near 0.5--0.6 $R_{\rm Sat}$,
i.e.~near the top of the demixing region, is required by Saturn models that aim at explaining the
$n=1,l=2,m=-2$ mode observed in Saturn's C-ring by a g-mode that is trapped in an extended,
inhomogeneous dilute core \citep{Fortney22}.
The dilute core regions of higher-$Z$ values in Jupiter and Saturn are deep enough 
that they do not matter for the determination of the He abundance at the top of the He-rain region, 
where $Z$ is still low. The upper limit of the $M_p$--$Z_p$ relation derived from exoplanets 
with measured mass and radius suggest $Z$ up to 0.9 for Saturn-mass planets and $Z$ up to 0.5 for 
Jupiter mass planets \citep{Thorngren16}.

For all planets, we assume a constant-$Z$ value along the adiabat that is varied 
between 0 and 0.9. This $Z$ is water. The  $Z$-value acts as a dilution of the He/H-ratio 
and it can influence the $P$--$T$ profile. If water in the He-rain region is dissociated it 
contributes two H-atoms per water molecule. This dilutes the He/H-ratio. If the adiabat is 
cold enough, water could also be in the superionic phase, where two protons are released per 
H$_2$O molecule, although if H/He phase separation occurs in a superionic environment is unknown. 
A low $Z$-value means that $Z$ can be rocks or that water has a low abundance.
We neglect the possibility that, if Uranus and Neptune are rock-rich, the dissociation of 
Fe-bearing rocks can lead to the absorption of H atoms into Fe-H alloys \citep{Horn23}. 
This process would tend to enhance (anti-dilute) the deep He/H ratio. However, we expect the 
number of H-atoms entering this pathway to be negligible.

Including heavy elements in the computation of the entropy affects the 
resulting adiabatic $P$--$T$ profile when $Z\gtrsim 0.2$ \citep{Baraffe08}. Volatiles 
in the entropy computation lead to colder adiabats. For example,  the icy Uranus model 
\citep{Bethkenhagen17}, with its purely icy interior ($Z_2=$1) has low core temperature of only 4000 K 
compared to the $\sim$6000 K of adiabatic Uranus models with a deep H/He-content of $\sim$0.1
\citep{Nettelmann13}. 
For simplicity however, our $P$--$T$ profiles are based solely on the EOS of H and He 
when using the CMS19 or the CD21 EOS, see Section \ref{sec:eos}. 

Given interior abundances in Z and Y, the adiabat is uniquely defined by the 1-bar surface 
temperature. Current Uranus and Neptune have $T_{\rm 1bar} \sim$ 75 K \citep{Lindal92}, but we calculate 
series' of adiabats where $T_{\rm 1bar}$ starts at high values corresponding to younger or more massive 
or closer-in planets.

\subsection{Equations of state}\label{sec:eos}

With the advent of extensive computer simulations an increasing number of H/He-EOS variants has
been offered over the past two decades that are in good agreement with results from compression 
experiments \citep{Desjarlais03}. Here, we use three different H/He-EOS: CMS19 \citep{Chabrier19}, 
CD21 \citep{Chabrier21}, and REOS.3 \citep{Becker14}. When $T_{\rm 1bar} < 128$ K, we use 
H/He-REOS.3 for the outer region up to 140 K. This choice led to a smoother transition in 
$Y_{\rm atm}(\Tone)$ than a cut at 100 K would. For the $Z$-component we use H2O-REOS, but only in 
conjunction with REOS.3.

Both CMS19 and CD21 EOS are convenient to use because they provide tabulated entropy data. CMS19-EOS 
provide tables for pure H and pure He for temperatures $T\geq 100\:$K. These tables can be combined 
using the linear mixing assumption. This way non-ideal effects on the entropy of mixing between H and He are 
ignored. The effective H-EOS of the CD21-EOS tables on the other hand does include non-ideal mixing effects. 
For CD21 EOS and the particular He mass fraction of 0.245, the entropy of the linearly mixed system 
will be the same as that of the full H-He mixture according to the binary mixture simulations of \citet{MH13}. 
One can therefore expect the resulting $Y_{\rm atm}$--$\Tone$-curve of intermediate non-ideal mixtures 
between $Y=0$ and 0.245 to fall in between the results for CMS19 and CD21 EOS. 
\citet{Howard23smix} have constructed non-ideal entropy corrections for any intermediate mixture. 
For Jupiter and its observed abundance of 0.238, the influence on interior $P$--$\rho$ profiles between 
the corrected version and CD21 EOS  is small. The smallness shows up for instance as an only 1~K difference 
in the 10--20 K shift of the $T_{\rm 1bar}$ temperatures that would be needed to lift the atmospheric-$Z$ 
from interior models up to $\sim 1.3\times$ solar in Jupiter. While $P$--$\rho$ is important for gravity, 
$P$--$T$ is important for demixing.

For H/He-REOS, we use our usual procedure of thermodynamic integration to compute adiabats for a 
given mixture EOS including water. This procedure requires only density and internal energy as a function
of $(P,T)$. In the mixture EOS, water is included by its contribution to both density and internal energy 
assuming linear mixing. Thus the contribution of the $Z$-EOS to the ideal entropy of mixing is implicitly
included in this case.

\begin{center}
\begin{figure}
\rotatebox{270}{\includegraphics[width=0.68\textwidth]{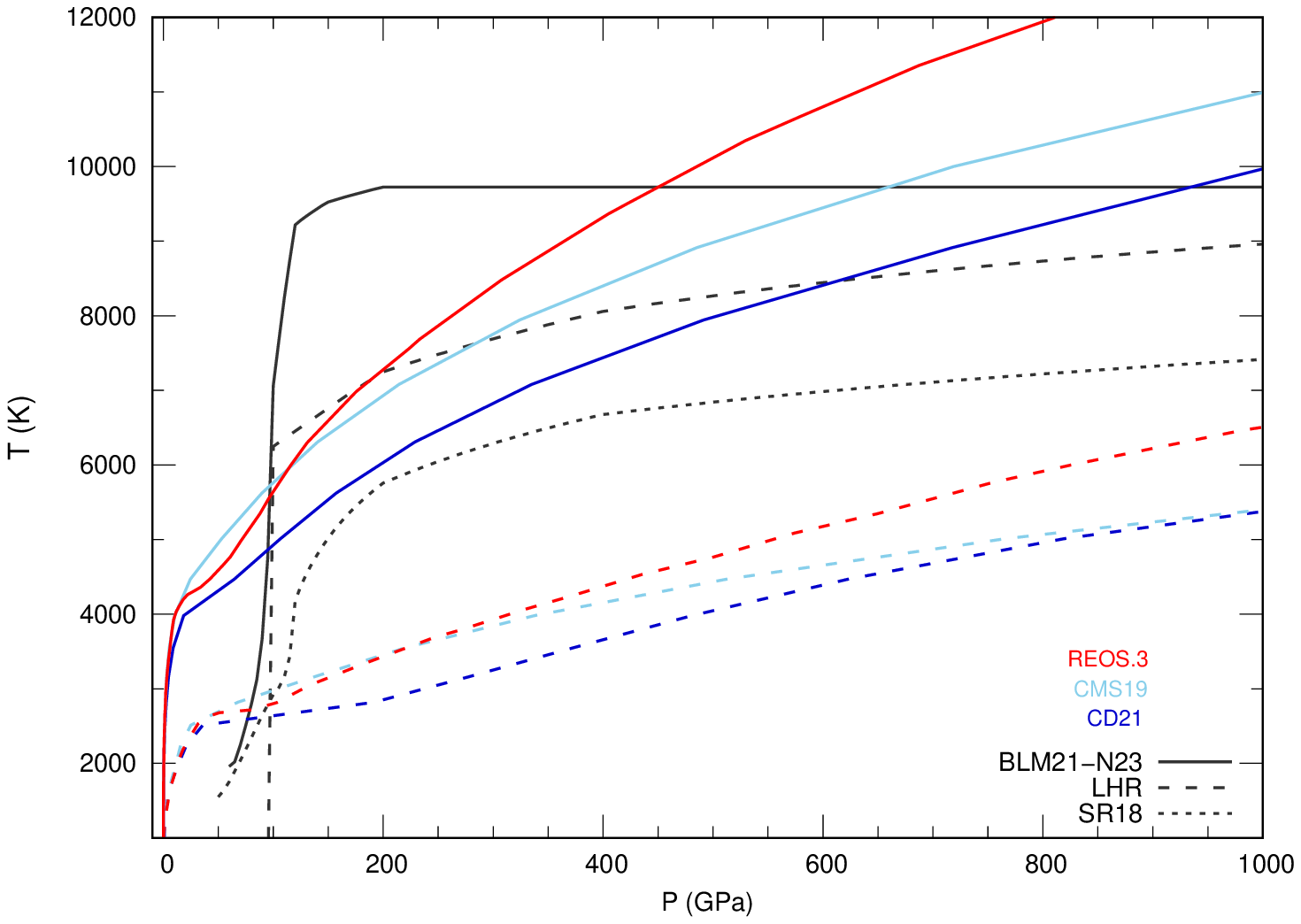}}
\caption{\label{fig:demixad}
Demixing curves and adiabats for $Y=0.27$. Grey lines are the demixing curves for the three H/He phase diagrams 
(solid: BLM21-N23, dashed: LHR0911, dotted: SR18). Colored lines show the adiabats constrained by 
$\Tone=166.1$K (solid) and $\Tone=75$ K (dashed) for the three H/He EOSs (red: REOS, blue: CD21, 
light blue: CMS19).}  
\end{figure}
\end{center}

Figure \ref{fig:demixad} shows H/He adiabats and demixing curves in  $P$--$T$ space for the three H/He-EOS 
and the three H/He-phase diagrams, respectively. All curves are for $Y=0.27$. The adiabats have either
 $\Tone=166.1\:$K as measured by the Galileo entry probe in Jupiter, or $\Tone=75\:$K representative 
of Uranus and Neptune. While all three phase diagrams suggest that He in the entire deep interior 
underneath the 1 Mbar level would demix in the ice giants, the predictions for  jovian planets are 
highly sensitive to both the phase diagram and the H/He-EOS.  For instance, only the CD21-adiabat is cold 
enough to touch the SR18 H/He-demixing region within the uncertainty of $\pm 500$ K of the latter. 
The rather warm REOS-adiabat suggests a demixing region in Jupiter between 1 and 4 Mbars for the 
BLM21-N23 H/He phase diagram, although the bottom pressure could be different if the He-gradient, 
mass conservation, and super-adiabaticity were taken into account as in \citep{Nettelmann15}, or if the 
demixing curve would flatten less strongly.

\subsection{Helium depletion in the planet}

At a given $T_{\rm 1bar}$, the adiabat $T(P; Y)$ is compared to the phase boundary $T_{\rm dmx}(P; Y)$.
When overlap is detected, we can record the pressure range over which demixing occurs; it is bounded by
$P_{\rm in}$ and $P_{\rm out}$.
An iterative procedure is employed that lowers $Y$ until the then colder adiabat (the adiabatic gradient
increases with $Y$) only touches the H/He phase boundary ($P_{\rm touch}=P_{\rm in}=P_{\rm out}$).
With the LHR0911 phase diagram, $P_{\rm touch} \sim 100$ GPa near the metallization pressure,
while with SR18, most solutions have $P_{\rm touch} \sim$ 200 GPa, which in turn is characteristic of the
metallization pressure in the vdw-DF1-based hydrogen system, see Section \ref{sec:HHeDiagram}.

This converged new helium abundance $Y_A$ satisfies $T(P_{\rm touch}; Y_A)=T_{\rm dmx}(P_{\rm touch}; Y_A)$.
It is the minimum equilibrium abundance. At even lower values $Y<Y_A$ along  the adiabat, the system would be
miscible. $Y_A$ is the helium mass fraction that is equivalent to the equilibrium particle fraction $x_A$, see Section \ref{sec:xA}.

At deeper levels below $P_{\rm touch}$, the equilibrium He abundance would increase with depth
until either the adiabat leaves the He-rain region at a sub-saturated level (Jupiter case)
or the remaining volume becomes too small to dissolve all the rained-out helium and He-layer formation occurs
(saturated Saturn case). For the dichotomy between Jupiter's and Saturn's internal structure due to
He rain, see \citet{MF20}. Here we ignore what happens to the deeper interiors.

\subsubsection{He depletion in the atmosphere}

If convective motions transport a He-poor parcel of concentration $x_A$ upward, it will act to deplete 
the region above when it dissolves. If convective motions transport a parcel of concentration
$x_{\rm He}>x_A$ downward into the demixing region, it will undergo demixing to this equilibrium abundance $x_A$.
If the convective overturn timescale is much shorter than the timescale of thermal cooling of the
entire planet, a steady state can be reached where the entire region above the He-rain region
will attain the equilibrium abundance $x_A$. In the adiabatic, convective case, we therefore expect
to the see this equilibrium abundance in the atmosphere and set $Y_{\rm atm}=Y_A$.
This procedure is the same as in \citep{Nettelmann15}.

\subsubsection{Atmospheric He abundance with TBL} \label{sec:methTBL}

Adiabatic, convective models fail to explain the observed luminosities of Uranus and Neptune \citep{Scheibe19}.
Deviations from the adiabatic, convective baseline case $Y_{\rm atm}=Y_A$ are therefore expected.

\begin{figure}
\begin{center}
\includegraphics[width=0.4\linewidth]{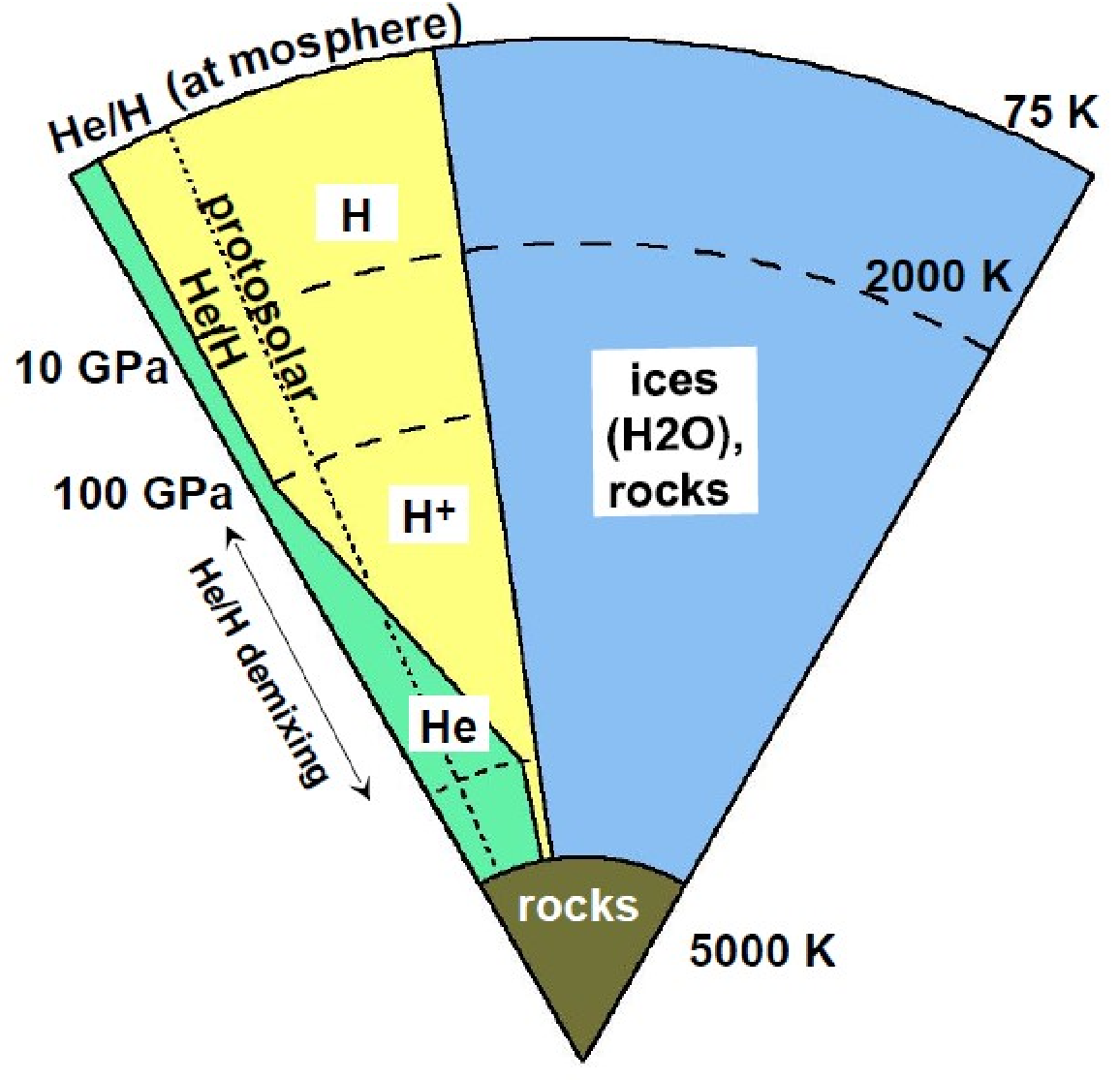}
\includegraphics[width=0.42\linewidth]{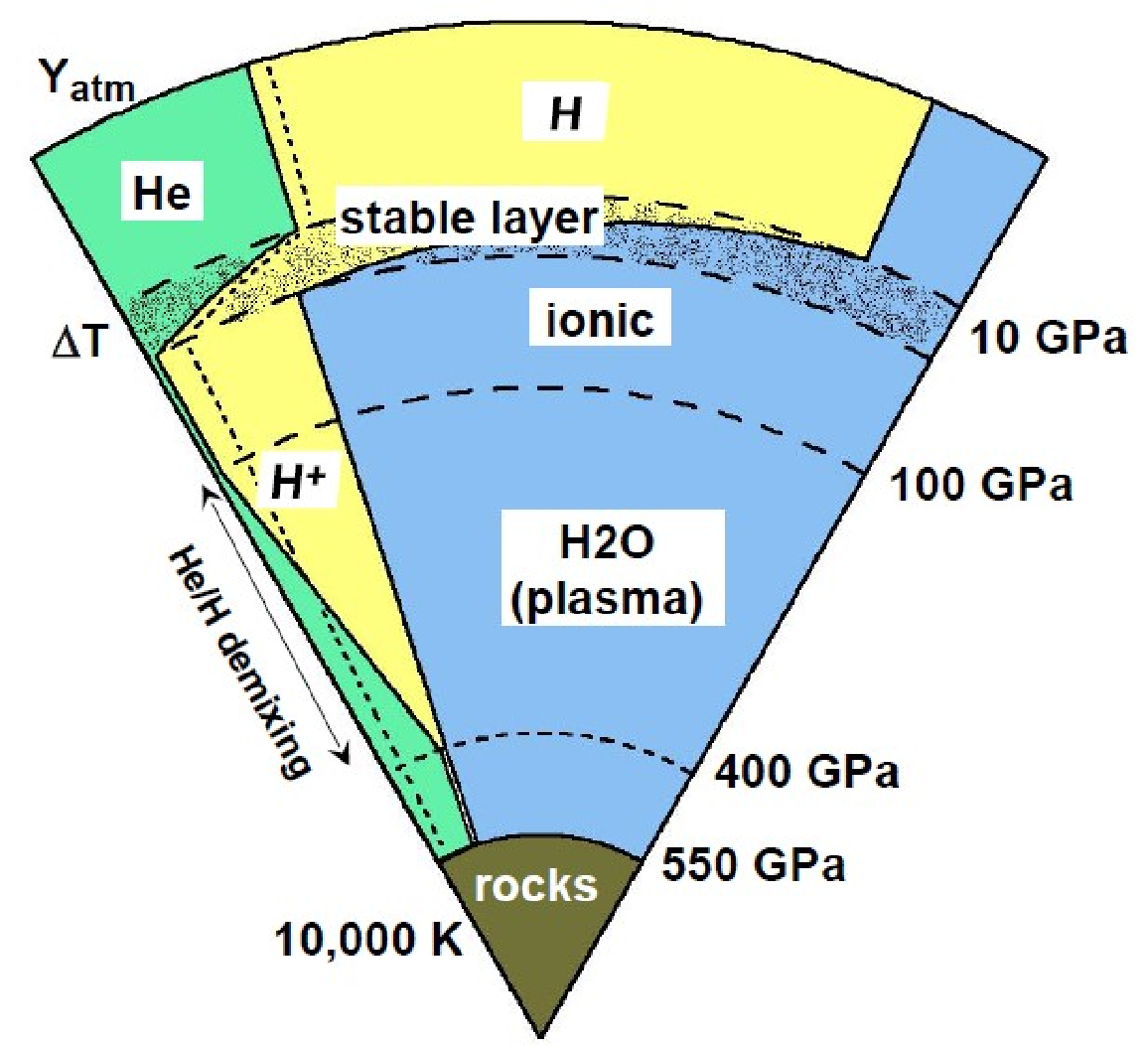}
\caption{\label{fig:tortenU}
Schematic 2D interior structure models of a Uranus or Neptune-like ice giant illustrating the mass 
fractions of hydrogen (yellow), helium (green), and water (blue) over radius (radial linear scale). 
At the center of the planet is a $\sim$1 $M_{\rm E}$ rocky core. Left: This model assumes that the 
depletion in He caused by H/He demixing at Mbar pressures extends all the way to the atmosphere 
due to convection or other means of efficient communication. Due to convection the $Z$-distribution 
should be homogeneous, which however is not supported by the gravity data. The dotted line shows a constant
protosolar He/H mass ratio of 0.27. Right: A stable layer (shaded) between the ice-poor outer and 
the ice/rock rich deep interior acts as a thermal boundary layer. It reduces the efficiency of heat 
and particle transport between the atmosphere and the deep interior. Primordial heat can be trapped 
inside, resulting in a temperature increment $\Delta T$. In addition, the stabilizing $Z$-gradient 
may permit a destabilizing He-gradient, so that the He depletion seen in the atmosphere could be weaker 
than at Mbar level.
}
\end{center}
\end{figure}

In another series of models we assume that a thermal boundary layer (TBL) exists between 
the $Z$-poor outer region and the $Z$-rich inner region (see Figure \ref{fig:tortenU}, right panel). 
In Uranus and Neptune, this transition occurs in the $\sim 10$ GPa region (see Sec.~\ref{sec:methProfiles}).
The idea is that if both envelopes were connected by convective motions, they would have erased the $Z$-gradient 
over time (see Figure \ref{fig:tortenU}, left panel). A $Z$-gradient itself inhibits convection unless 
it is accompanied by a sufficiently strong, de-stabilizing temperature gradient. However, the initial 
heat budget deep in Uranus, and presumably also Neptune, does not seem high enough to wipe out its entire 
$Z$-gradient \citep{Vazan20}. If the $Z$-gradient inhibits convection, it also reduces the transport of heat. 
This is a TBL.

A TBL influences the thermal evolution.
Conversely, as the outer convective region cools efficiently while the region interior to the stabilizing
$Z$-gradient cannot, a super-adiabatic temperature gradient builds up over time across the TBL.
Once the temperature gradient is strong enough, either convective mixing sets in \citep{Vazan15} or heat
diffusion becomes efficient enough to release heat from the deep interior. In the latter case it has been
shown that Uranus and Neptune could appear initially fainter than a planet without a TBL, while after
some time they could appear brighter \citep{Scheibe21}. Thermal evolution models easily place Neptune
on the bright side today, whereas due to the stronger incident flux received by Uranus, thermal evolution models
that produce effective temperatures close to the observed one of $59 \pm 1$ K suggest that its outer envelope
entered a state of thermal equilibrium with the incident flux quite some time ago, which dramatically slows
down the cooling of the deep interior \citep{Scheibe21}. Models that assume that the barrier to convection
existed since early on predict that most of the primordial heat is still conserved today. Such models thus have
a strong temperature increment of up to $\sim 8000\,$K across the TBL, which is stabilized by a strong 
compositional gradient, except perhaps toward the very top of the TBL where the temperature is low and
the luminosity high \citep{Scheibe21}. One could also imagine that the $Z$-gradient
formed later in the evolution, perhaps as a result of H$_2$/H$_2$O phase separation \citep{Bailey21}.
In that case, the present temperature increment could amount to only a few 100 to few 1000 K \citep{Scheibe21}.

From a structure models perspective, \citet{Neuenschwander24} quantify the temperature increment by comparing 
their piecewise polytropic profiles to equations of state.
If the condition of constant entropy at homogeneous composition is violated, they determine the minimum
increment in $Z$ and the depending maximum increment in $T$ that would keep the layer locally Ledoux-stable.
Unsurprisingly they arrive at rather large temperature increments with a minimum and mean
of respectively $\Delta T=$3000 K and 8000 K in the 0.6--0.8 $R_U$ region where this effect is most pronounced.

In this work we represent a TBL by a sharp temperature jump at 10 GPa. While guided by interior models
that fit the gravity data, this choice matters. The deeper the assumed TBL is located, the less sensitive
to thermal effects will be the EOS and the adiabat. Stronger temperature increments could therefore be possible 
for deeper assumed TBLs. However, we expect a variation within the  10 to 20 GPa range to have a minor effect,
and therefore, the fixed mass-shell location at 13 $\ME$ as assumed in \citet{Scheibe21}, 
which for present Uranus is at $\sim$20 GPa, not to be responsible for the much larger $\Delta T$ values found
to be necessary to explain the luminosity in some of their evolution models than if they had chosen 10 GPa. 

We vary $\Delta T$ between 0 and the maximum where the deep interior becomes too 
warm for H/He demixing to occur. The effect of the TBL on $Y_{\rm atm}$ is only through the warmer deep interior. 
We do not model the reduced particle transport across the TBL. While this is an obvious inconsistency, 
it allows us to estimate the minimum $Y_{\rm atm}$ as a function of $\Delta T$ given its location.

\section{Observed Atmospheric Helium Abundances}\label{sec:obsHHe}

For all four outer planets there are data from the combined analysis of Voyager occultation and IR
remote sensing \citep{Gautier81,Conrath84,Conrath87,Conrath91} as well as values from various
independent sources.  What has been declared the \textit{final} Voyager values is shown by grey
symbols in Figures \ref{fig:YatmT_LHR}---\ref{fig:YatmT_BLM}; revised and other observed He abundances 
are shown as black symbols.

\paragraph{Jupiter: observations}
The Galileo entry probe value of Y=$0.238 \pm 0.005$ \citep{Zahn98} suggests that some He rain did occur. 
As both the mass spectrometer and the He abundance detector (HAD) aboard the Galileo entry
probe yielded consistent results \citep{Niemann98,Zahn98}, these entry-probe measurements are generally
considered reliable. In Figures \ref{fig:YatmT_LHR}--\ref{fig:YatmT_BLM} we plot the Galileo/HAD 
measurement of $Y_{\rm atm}$ and $\Tone$ ($166.1 \pm 0.8\,$K) as a dark-red dot. As the 
in-situ measurements were made in a hot spot, this temperature however may not be representative of 
Jupiter's global $T_{\rm 1bar}$; moreover, Voyager I observations indicated a latitudinal variation 
in $\Tone$ by $\sim$7~K across the equator.
The Voyager data have been reanalyzed recently \citep{Gupta22}, which results in a current $T_{\rm 1bar}$ 
estimate of 164---174 K; this is $+4\:$K higher than the original Voyager estimate \citep{Lindal92}.
A re-analysis was kind of overdue as temperatures inferred from radio occultations depend on the
composition of the atmosphere, and the helium abundance used for the Voyger occultation analysis was 
lower than the later Galileo value.

\paragraph{Saturn: observations}
Despite several efforts to determine Saturn's atmospheric He abundance, we cannot tell which one, if any,
of the measured values reflect Saturn's true He abundance. The measurements include a mole fraction of
$q_{\rm H_2} = 0.90\pm 0.03$ from Pioneer II occultation and IR remote sensing \citep{OrtIng80}, $Y_{\rm atm}=$0.18--0.25 from Voyager IRIS only \citep{CG00}, $Y_{\rm atm}=$0.158--0.217 
from stellar occultations and Cassini UV remote sensing \citep{Koskinen18}, ${\rm He/H}_2 =$ 0.10--0.16 
from the INMS mass spectrometer at Cassini Grand Finale Tour \citep{Waite18}, and $Y_{\rm atm}=$0.075--0.130
from Cassini CIRS \citep{Achterberg20}.

Where the ratio $q_{\rm He}/q_{\rm H_2}$, short He/H$_2$, is provided, we convert it to $Y_{\rm atm}$ using
\begin{equation}\label{eq:qHeqH}
	Y_{\rm atm} = \frac{m_{\rm He}\:{\rm He/H_2}}{m_{\rm He}{\rm He/H_2} + m_{\rm H_2}}
\end{equation}
with $m_{\rm He}=4.0026$ and $m_{\rm H_2}=2.0158$ \citep{Conrath87}. Where the mole fraction $q_{\rm He}$
is provided without clear indication of the mole fractions of other species like methane or $N_2$,
we convert according to
\begin{equation}\label{eq:qHe}
	Y_{\rm atm} = \frac{m_{\rm He}q_{\rm He}}{m_{\rm He}q_{\rm He} + m_{\rm H_2}(1-q_{\rm He})}
\end{equation}
with $q_{\rm He}+q_{H_2}=1$.

\paragraph{Uranus and Neptune: observations}

Early analysis of the Voyager data suggested that the atmospheres of Uranus \citep{Conrath87} and
Neptune \citep{Conrath91} have preserved the protosolar value, with perhaps a slight enhancement in
Neptune \citep{Atreya20}. However, the He abundance from occultation data is inferred from the
mean molecular weight, and thus different abundances of methane or other atmospheric constituents
would result in a different inferred helium abundance.
\citet{Sromovsky11} found that IR spectra taken by Voyager 2 and near-IR spectra taken
by HST are better explained by a 2.3--4\% methane volume mixing ratio above a thick methane cloud
deck compared to its neglection in the \textit{final} analysis of the occultation data by \citep{Conrath87}.
According to \citet{Sromovsky11} the Voyager refractivity profiles would therefore be better
explained by a lower He abundance implying a moderate depletion for Uranus. We convert the He/H$_2$ mixing 
ratios of their models D1, F1, D, to He mass abundance using Eq.\ref{eq:qHeqH}.

A similar effect is seen for Neptune, where inclusion of 0.3\% N$_2$ shifts the He abundance downward
to around the protosolar level \citep{Atreya20}.

\section{Results for the Atmospheric Helium Abundance}\label{sec:results}

The 1-bar temperature serves the interior models as an outer boundary condition for the adiabat.
The higher $T_{\rm 1bar}$, the warmer the adiabatic interior. For high enough $T_{\rm 1bar}$
a protosolar abundance of helium will remain fully miscible in the hydrogen-environment. This means
that we expect the atmosphere of a warm planet to be of protosolar H/He ratio, $Y_{\rm atm} = Y_{\rm proto}$
unless other processes than H/He immiscibility act to modify the atmospheric H/He ratio.
The lower $T_{\rm 1bar}$, the colder the adiabat. With decreasing $T_{\rm 1bar}$, a planetary adiabat
will eventually begin to overlap with the H/He demixing diagram. The colder the adiabat, the lower is
the equilibrium abundance $Y_A$.
According to this simple model, we  expect to see a stronger atmospheric helium depletion with respect
to the protosolar value the lower the $T_{\rm 1bar}$ of the planet.

\begin{figure}
\begin{center}
\rotatebox{270}{\includegraphics[width=0.68\textwidth]{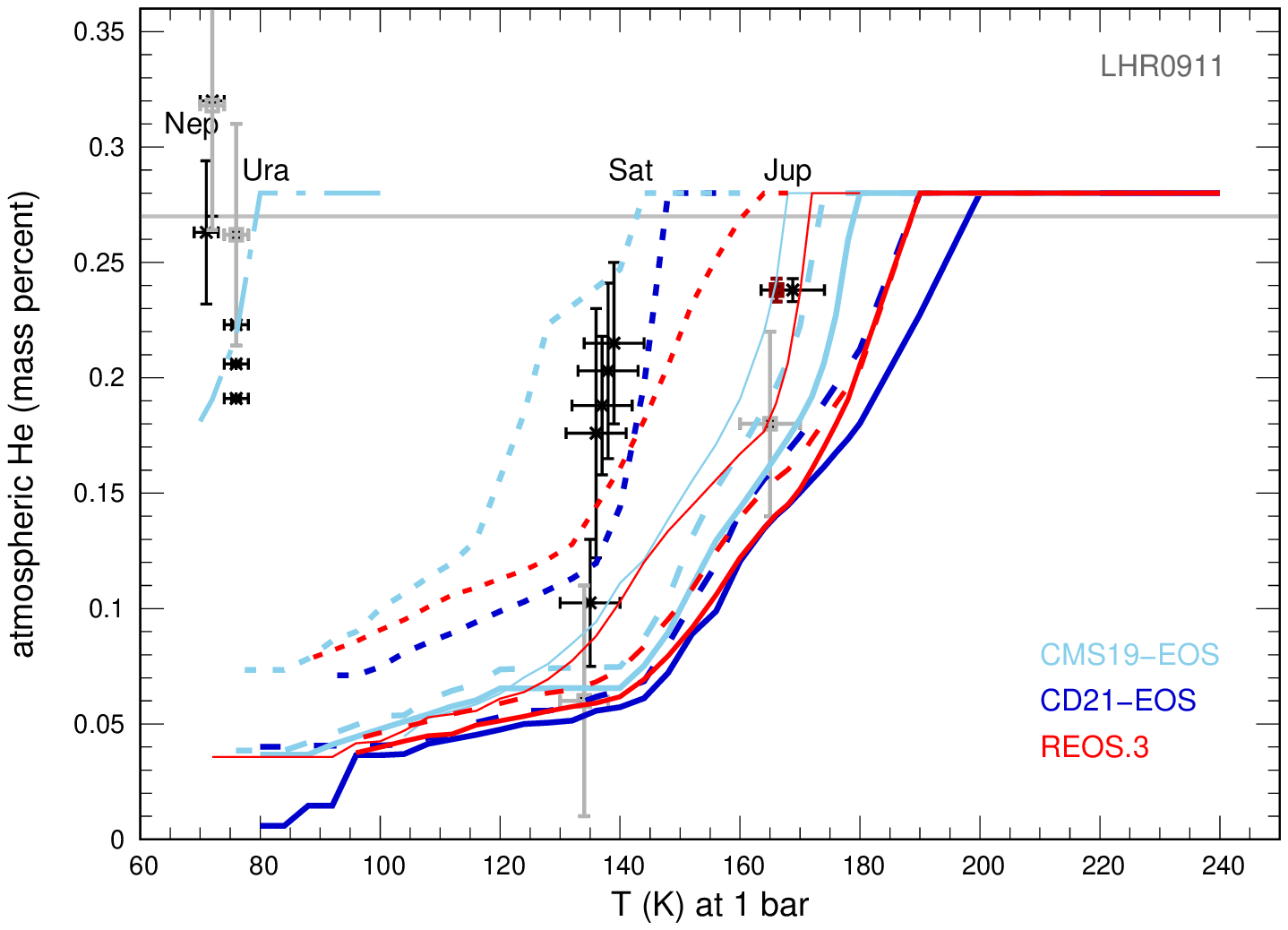}}
\caption{\label{fig:YatmT_LHR}
Atmospheric He abundances in the giant planets from observations (grey, black symbols 
with error bars) and predicted (lines) by the BLM21-N23 H/He phase diagram and the three different H/He-EOS
REOS (red), CD21 (blue), CMS19 (light blue). 
Symbols: grey symbols are the \textit{final} Voyager data for Jupiter and Saturn
\citep{Conrath84}, Uranus \citep{Conrath87}, and Neptune \citep{Conrath91} displayed at the Voyager
1-bar temperatures of \citet{Lindal92}. Black dots for Saturn and Neptune are slightly shifted in 
$T_{\rm 1bar}$ for readability.
Black symbol for Neptune: revised Voyager with 0.3\% $N_2$ \citep{Atreya20};
black symbols for Uranus: models D1, F1, G with enhanced methane abundances \citep{Sromovsky11};
Lines: Thick lines are for the $Z_{H2O}=0$ (solid), 0.5 (long dashed) and 0.9 (dotted). 
Thin solid lines are for constant shifts in $T_{\rm dmx}$ chosen to reproduce 
the Galileo entry probe value of Jupiter, here $dT_{ \rm dmx}=-400(420)\:$K for CMS19(REOS).
Dashed-dotted lines are for ice giant models with a thin thermal boundary layer at 10 GPa 
across which the temperature increases, here by $\Delta T=2000\:$K (CMS19), and $Z=0$. 
Grey horizontal line indicates protosolar value of $Y_{proto}=0.27$.}
\end{center}
\end{figure}

\begin{figure}
\begin{center}
\rotatebox{270}{\includegraphics[width=0.68\textwidth]{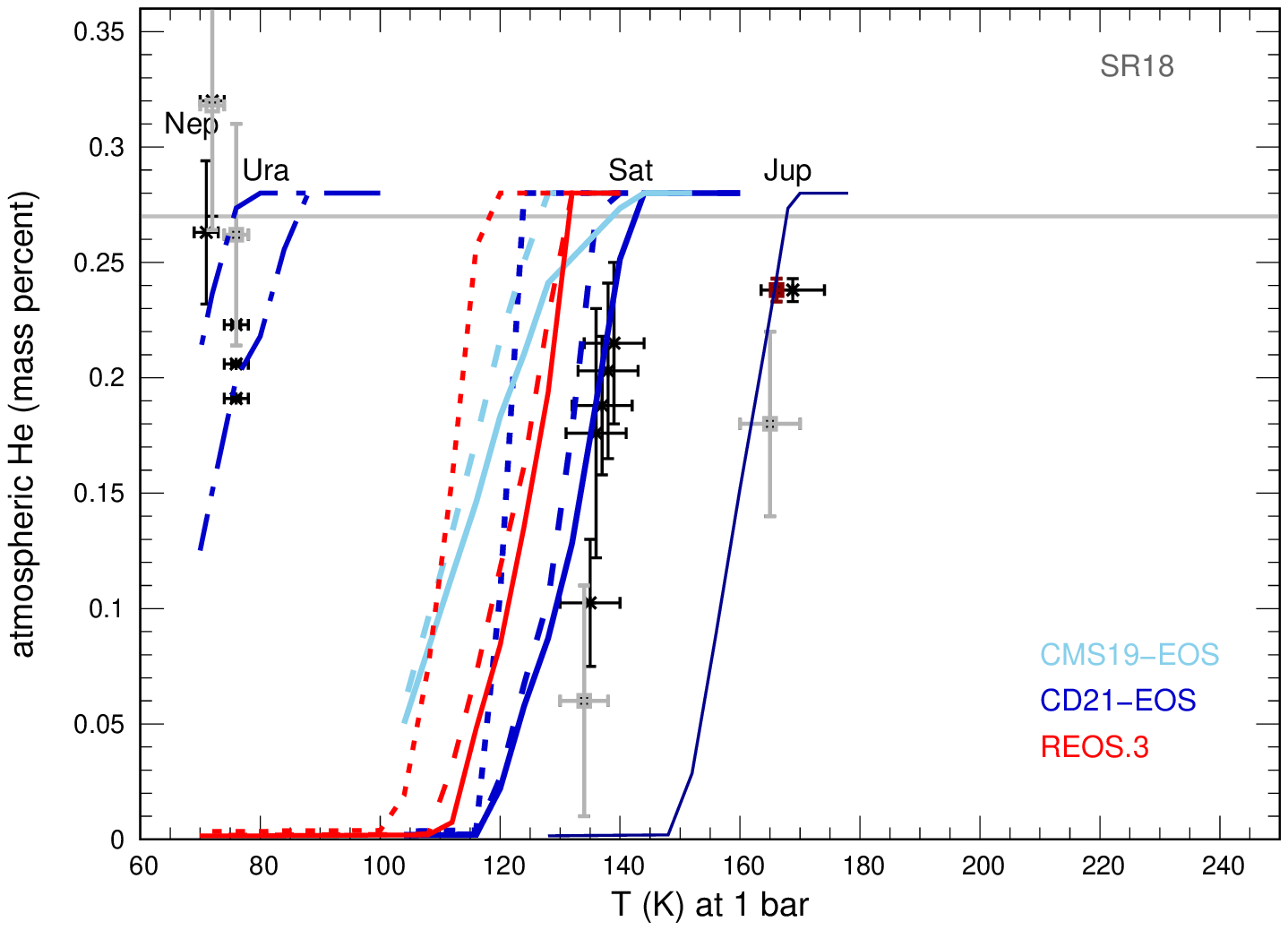}}
\caption{ \label{fig:YatmT_SR}
Same as Fig. \ref{fig:YatmT_LHR} but for the SR18 H/He phase diagram and a TBL with  $\Delta T$=1300 or 
1500~K (CD21) and a phase diagram shift of +1000$\:$K (CD21).}
\end{center}
\end{figure}

\begin{figure}
\begin{center}
\rotatebox{270}{\includegraphics[width=0.68\textwidth]{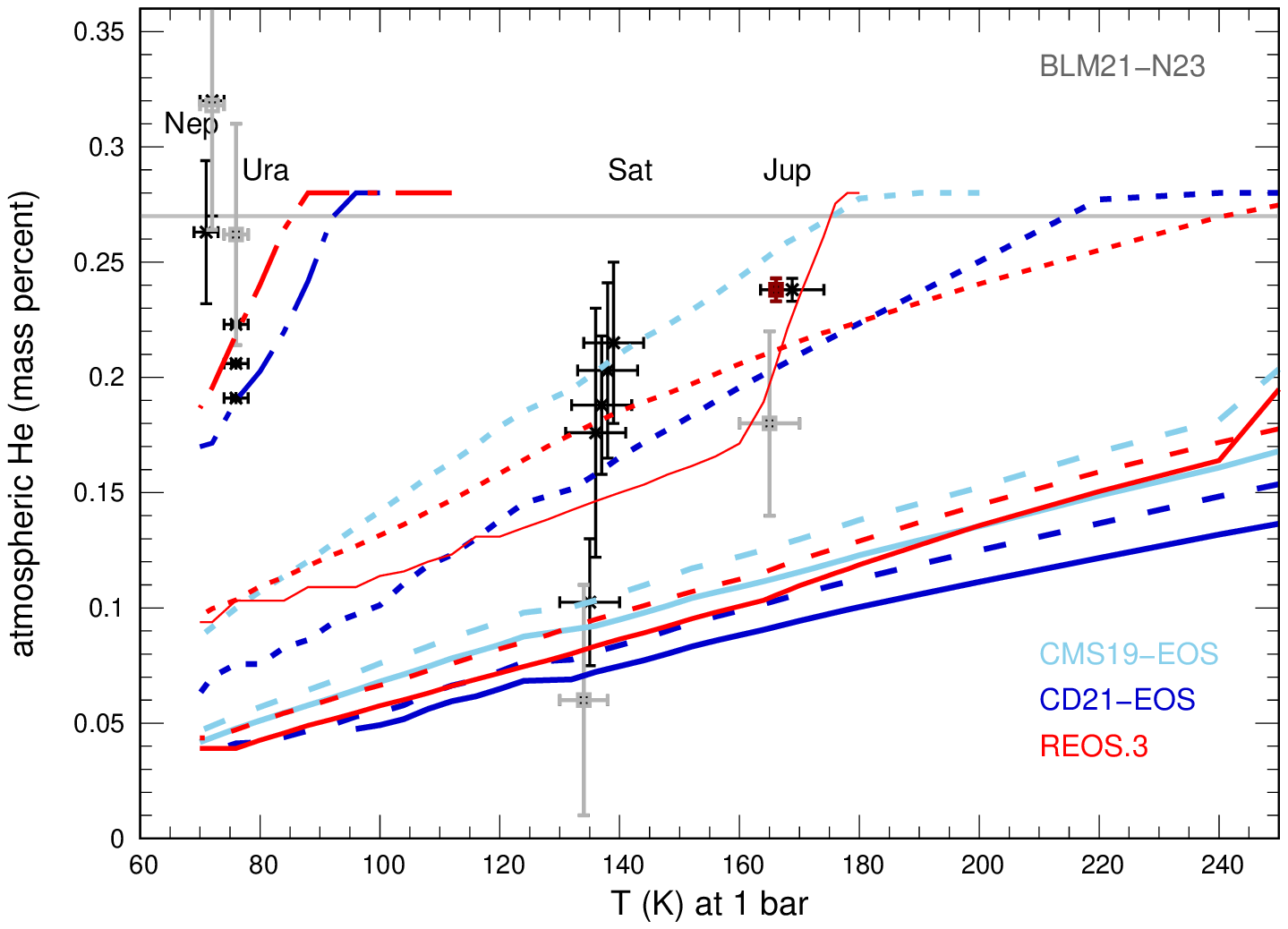}}
\caption{ \label{fig:YatmT_BLM}
Same as Fig. \ref{fig:YatmT_LHR} but for the BLM21-N23 H/He phase diagram and a TBL with $\Delta T=3500\:$K 
(CD21, REOS.3) and a phase diagram shift of -2820$\:$K (REOS.3).}
\end{center}
\end{figure}

In Figures \ref{fig:YatmT_LHR}---\ref{fig:YatmT_BLM} we compare our predictions of the atmospheric 
helium depletion due to H/He phase separation at depth to the observed atmospheric He abundances of 
the outer planets.

Overall we note a strong influence of the phase diagram and of the H/He EOS used.
CD21-EOS yields the coldest adiabats and therefore the strongest depletion. CMS-19 EOS yields 
the highest temperatures at 1 Mbar, but not at 2 Mbars, where H/He-REOS is warmer. To reach 
the equilibrium concentration at the tangential $P_{\rm touch}$, which is typically somewhere 
in between, apparently a less strong depletion is needed than for REOS. This is an effect of 
the slopes of the adiabat and the demixing curves.  

Since the SR18 phase diagram has the lowest demixing temperature at protosolar He abundance, 
the initial entry into the demixing regions occurs at the lowest 1-bar temperatures. But since 
the temperature are rather low upon first entry and the isobaric demixing temperatures are rather 
flat with decreasing He abundance (Fig.~1), the equilibrium He abundance must fall steeply with 
decreasing T1bar. The opposite holds for the BLM21-data, where the entry-point occurs at high $\Tone$.
As a result, $Y_{\rm atm}$ falls slowly with $\Tone$ for BLM21-N23. 

The He/H dilution-effect with increasing Z can clearly be seen in form of a reduced He depletion, 
but this effect becomes relevant only for $Z>0.5$. For $Z=0.9$, the REOS-adiabats show stronger 
depletion than the CD21-adiabats. This is the effect of the temperature-influence of $Z$
on the ideal-entropy of mixing to make the adiabat colder, and since this effect is included only for
REOS-adiabats, the REOS-adiabat becomes colder while the CD21 adiabat does not at high-Z. 
Colder adiabats lead to stronger depletion. 

\paragraph{Jupiter: predictions}

Given the uncertainties in the H/He phase diagram, in the H/He EOS, in Jupiter's internal structure,
and our simplifying assumption of an adiabatic interior, it is not to be expected that
the Galileo value would right-away  be reproduced by our atmospheric helium depletion model.

For the SR18 phase diagram (Fig.~\ref{fig:YatmT_SR}), Jupiter seems to be too warm for demixing to occur 
regardless of the EOS. On the contrary for to BLM21-N23 (Fig.~\ref{fig:YatmT_BLM}) Jupiter 
would experience very strong depletion of $Y\sim$0.1 almost regardless of the H/He EOS and $Z<0.5$
With LHR0911, the depletion is around 0.15. 

Let us assume that some of the uncertainties in regard of the phase diagram and the EOS can be captured
by shifting the phase diagram in temperature to match the Galileo value. That procedure
has been adopted before \citep{Nettelmann15,MF20} to compute the influence of helium rain on the 
evolution of Jupiter. In a pioneering study, \citet{FH03} shifted the phase diagrams
of that time to reproduce the luminosity of Saturn.
For the LHR0911 H/He phase diagram only a moderate shift by $\sim 400\:$K is needed which is well 
within the uncertainty of $\pm 500\:$K of the phase diagram. Shifting the phase diagram vertically
in temperature leads to a horizontal shift in the $Y_{\rm atm}(\Tone)$ curves. Any combination that
reproduces the Galileo value predicts a rather strong He depletion for Saturn of 0--0.14.

\paragraph{Saturn: predictions}

The top of the demixing region in Saturn begins between 0.65 $R_{\rm Sat}$ (1 Mbar, LHR0911) 
and 0.55 $R_{\rm Sat}$ (2 Mbar, SR18), which is deeper inside than in Jupiter where 
the 1 Mbar region is at 0.86 $R_{\rm Jup}$ and 2 the Mbar level at 0.81 $R_{\rm Jup}$.

One may think that knowledge of Saturn's tropospheric He abundance in addition to that of Jupiter 
could help constrain the H/He phase diagram: a prediction that works for Jupiter should also work 
for Saturn if Saturn's outer envelope is similar to that of Jupiter.
Adjusted to Jupiter (thin solid lines) or not, both LHR0911 and the BLM21-N23 data predict 
$Y_{\rm atm}<0.14$ for Saturn. In contrast, the SR18 diagram with the CD21-EOS 
predicts moderate depletion of 0.15--0.25 for Saturn. 

In Section \ref{sec:discussGG} we discuss the inversion of what a measured value for Saturn 
by an entry probe could imply for the interior structure and the phase diagram. However we caution 
that further planet properties may influence the atmospheric helium abundance rendering the inversion 
to be non-unique. An important such property would be an outer stable layer above the 
He-rain zone (OSL). Such has been suggested for Jupiter as a reason for the slow-down of the zonal winds 
and to  be necessary to maintain zonal winds at high latitudes within the tangent cylinder \citep{Wulff22}.  
A stable layer can be sub-adiabatic or super-adiabatic. A sub-adiabatic OSL (subOSL) could be caused by a 
radiative zone due to alkali metal depletion \citep{Guillot94}, which indeed may be needed to explain 
Jupiter's low opacity in the $\sim$1 kbar region as observed by Juno \citep{Bhatta23}. A subOSL would 
lead to colder interiors and thus stronger He-rain. In this case the He depletion in the atmosphere 
would be weaker than in the Mbar region: a de-stabilizing inverted $Y$-gradient.
A super-adiabatic OSL (supOSL) is a TBL. It implies warmer interiors with weaker He depletion at depth. 
The He-abundance in the atmosphere can be higher than depth: a de-stabilizing inverted Y-gradient 
and be higher than in the adiabatic case.

To stabilize a super-adiabatic layer against convection an additional Z-gradient is required. This is 
consistent with current Saturn and ice giant structure and models but not with current Jupiter models. 
Thus a supOSL is an option for Saturn but not for Jupiter. An OSL offers options 
to bring both Jupiter and Saturn under one umbrella even with the cold SR18 phase diagram and the
cold CD21 adiabat, which otherwise seems difficult according to Figure \ref{fig:YatmT_SR}.

\paragraph{Uranus and Neptune: predictions}

Under the assumption of an adiabatic interior, the LHR0911 H/He phase diagram predicts a strong 
($Y_{\rm atm}< 0.1$) He depletion in the cold atmospheres of Uranus and Neptune.
If there is a TBL, the deep interior will be warmer, and helium depletion at depth will be seen with a delay
in the atmosphere due to  the finite diffusivities of He atoms. By neglecting this delay we overestimate
the He depletion. The respective dot-dashed curves in Figs.~\ref{fig:YatmT_LHR}--\ref{fig:YatmT_BLM} 
are thus lower limits  to $Y_{\rm atm}$. For $\Delta T> 4000\,$K the interior would become so warm that 
H/He-phase separation can not occur in any of the considered cases.

With the help of a TBL, moderate depletion can be obtained, in agreement with the revised Voyager
measurements of Uranus. However, the depletion is highly sensitive to $\Delta T$. As argued above, 
$\Delta T$ could be smaller with a stronger He depletion at depth but a delayed observable depletion 
in the atmosphere, but $\Delta T$ could also be larger for a deeper TBL.

Some of the evolution models that are adjusted to reproduce the luminosity of Uranus and Neptune
\citep{Scheibe21} and some structure models \citep{Neuenschwander24}, suggest that $\Delta T$ could also 
be substantially larger in which case no depletion at depth occurs and thus no depletion in the atmosphere.

This leaves us with (at least) two options for Neptune:
(i) Helium is present at depth but due to a strong TBL, H/He phase separation  does not occur;
(ii) Helium is not present at depth.

\section{Discussion}\label{sec:discussion}

\subsection{H/He phase diagram.}

It has become common to shift the phase diagram in temperature to reproduce the Galileo value under 
the assumption of an adiabatic interior \citep{FH03,Nettelmann15,MF20}.
That a modification of the LHR0911 data set is needed to match the Galileo value is not surprising, 
as that data set neglects non-ideal contributions to the entropy of mixing and therefore can come 
close to the real demixing behavior, which we stress is still unknown, only by coincidence. 
That the required shift in temperature of $\sim$400~K  (with CMS19 and REOS) is well within the 
uncertainty of $\pm 500$ K at 1 Mbar to $\pm 1000$ K at Mbar \citep{LHR09} of the data set itself 
suggests that such a coincidence could indeed be the case.

On the other hand, the Galileo probe value may not be right-away reproduced for other reasons.
For instance, Jupiter's interior may not be fully adiabatic. A stably stratified layer atop the demixing region
is required by magnetic field models \citep{Moore22} and in order to slow-down the zonal
flows \citep{Christensen20,Wulff22}.

Consider a H/He phase diagram that reproduces the Galileo value for an assumed adiabatic interior
while the real interior has a stable layer above the demixing region.
If super-adiabatic, the stable layer would then require an upward shift $\Delta T_{\rm dmx}>0$ 
of this phase diagram. If sub-adiabatic, the stable layer may require a downward shift
$\Delta T_{\rm dmx}<0$ or an upward shift. 
If the applied H/H phase diagram seems to over-estimate the atmospheric He abundances in Jupiter, like the SR18 one does, introduction of a subadiabatic OSL is the only way to reduce the atmospheric value. 
If the applied H/He phase diagram seems to under-estimate Jupiter's atmospheric value, both a subOSL or a supOSL can lift $Y_{\rm atm}$ upward. Quantitative assessments are needed.
Knowledge of the H/He phase diagram therefore provide valuable information on Jupiter's interior
structure. More experimental efforts are needed to constrain the H/He phase diagram.

\subsection{Gas Giant Internal Structure}\label{sec:discussGG}

Let us assume the H/He phase diagram is uncertain within  the range given by SR18 on the cold end
and BLM21-N23 on the warm end, and that the adiabatic $P$--$T$ profile is uncertain within CD21
on the cold end and CMS19 on the warm end around 1 Mbar and REOS at higher pressures.
We here consider several inverted cases to see what, if anything, a potential Saturn entry probe He abundance
measurement could teach us about the H/He phase diagram or the internal structures of Jupiter and Saturn.
We consider the two extreme cases. Case (1) is a warm BLM21-N23 like phase diagram and cold CD21-like adiabat,
which leads to strong He depletion inside Jupiter.  Case (2) is a cold SR18-like phase diagram and a warm adiabat,
which leads to no demixing in Jupiter. Suppose a Saturn entry probe measures:
\begin{itemize}
\item
\textit{He/H in Saturn equal to Jupiter in case (1)}
Jupiter in this case must\footnote{The word 'must' abbreviates 'may have, among alternative explanations which
escaped our perception'} have a SSL atop the demixing region. Also Saturn must have a SSL and it must 
be less permissible than in Jupiter. What is seen in the atmosphere reflects the properties of the SSL. 
These properties must be able to explain the subsolar Ne/He in Jupiter's atmosphere.
Knowing Ne/He for Saturn would place additional constraints on the SSL.
\item
\textit{He/H in Saturn much less than in Jupiter in case (1)}
Jupiter must have a SSL. Saturn may have a Jupiter-like SSL. The lower He/H in Saturn's atmosphere occurs 
because He-rain in Saturn begun earlier in the evolution so He atoms had more time to travel through the SSL. 
\item
\textit{He/H in Saturn higher than in Jupiter in case (1)}
Jupiter must have a SSL. Saturn must have a thick SSL. While a thick SSL is indicated by the seismic data, 
that SSL would occur deeper inside underneath the He-rain region \citep{MaFu21}. If sub-adiabatic, 
it acts to shield 
the atmosphere from the He-depletion at depth. If super-adiabatic it acts to keep the interior so warm 
that He-rain does not or if so occurs only weakly in Saturn. Indeed, \citet{LC13} have suggested 
that a permanent thick SSL with a Z-gradient could have prevented Saturn's deep interior from 
efficient cooling so that today, Saturn's excess luminosity would result from primordial heat 
floating along a superadiabatic gradient rather than from the gravitational energy released by 
falling He droplets. This is a cold Jupiter and warm Saturn case.
\item
\textit{He/H in Saturn much less than in Jupiter in case (2)}
Jupiter must have a subSSL because of the phase diagram. The sub-adiabaticity in this case is 
mandatory in order to have a cold enough deep adiabat for He-rain to occur. An SSL is not needed for Saturn.
\item
\textit{He/H in Saturn higher than in Jupiter in case (2)}
Jupiter must have a subSSL because of the H/He phase diagram, Saturn as well. 
\end{itemize}
None of the considered cases excludes a particular phase diagram.

\subsection{Ice Giant Internal Structure}\label{sec:discussIG}

Let as assume we knew the H/He phase diagram and the H/He-EOS precisely. This leaves us with several
options of what to expect for the atmospheric helium abundance of the ice giants depending on their
internal structure. We list some cases below from the point of view of a putative probe-measurement.

\begin{itemize}
\item
\textit{Protosolar H/He.}
If there is no helium deep inside, H/He phase separation does not take place, and therefore we would expect
to see the protosolar value in the atmosphere unless there are other chemical processes that affect hydrogen
and helium differently. Observations indicate a protosolar H/He ratio in the  atmosphere of Neptune \citep{Atreya20}.
The currently rather large uncertainty in the low-order gravitational harmonics of Neptune permits
a rather wide range of internal compositions. One of such compositions is a pure water, H/He-free deep interior
if the entire planet is assumed to be adiabatic \citep{Nettelmann13}, or with some admixture of rocks if it
is warmer \citep{Nettelmann16}. Thus a He-free deep interior is an option for Neptune.

We can also expect to see the protosolar ratio if there is helium deep inside and H/He phase separation does
take place, but if the particle exchange between the interior and the atmosphere is
strongly inhibited. An indication for a barrier to particle transport could be that the 
deep interior below a few GPa or a few tens of GPa is more enriched in heavy elements than the outer
region \citep{Nettelmann13} for a wide range of thermal profiles \citep{Podolak19}. A barrier to convection
in regard of inhibited heat transport is a possible solution to explain the low luminosity of Uranus
\citep{Nettelmann16,Scheibe21} and the high luminosity of Neptune \citep{Scheibe21}.

In addition, we would expect no atmospheric He depletion if the deep interior were much wrmer than along
an adiabat. Strong temperature increments $>3000$~K can explain the luminosity of both planets \citep{Scheibe21}
and are predicted by some polytropic structure models \citep{Neuenschwander24}.

In all of these cases one would expect a protosolar or slightly enhanced Ne/H ratio, as Ne will not 
be removed from the atmosphere by partitioning into falling He-droplets at depths \citep{Stevenson98}
but may be slightly enriched if the accreted nebular gas was H/He-depleted.

\item
\textit{Strong Atmospheric Helium depletion.}
Strong atmospheric helium depletion in Uranus and Neptune is predicted by any of the considered
H/He phase diagrams and their modifications, simply because their surface temperatures of $\sim 75$ K
are so much lower than those of Saturn ($\sim 135\,$K) and Jupiter ($\sim 170\,$K); if the gas giants 
experience H/He phase separation, so would planets with colder adiabats even more. This scenario implies 
that the interiors of Uranus and Neptune are adiabatic and do exchange particles with the atmosphere, 
and that helium is present at depths of Mbars. We showed that even for high ice mass fractions of up to 90\%, 
consistent with structure models, He-rain would strongly deplete the region above to levels 
of $Y_{\rm atm}\ll 0.1$

In this case one would also expect to see a strong Ne/H depletion, as Ne will be removed by partitioning
into falling He droplets.

\item
\textit{Weak Atmospheric Helium depletion.}
On Uranus, a weak atmospheric helium depletion is indicated by the observed mean molecular weight
profile inferred from the Voyager occultation experiment if an enhanced methane abundance is assumed as
would best explain the IR and near-IR spectra \citep{Sromovsky11}. Our model of H/He phase separation
predicts weak helium depletion if the (shifted) LHR0911 data are used and most importantly, if a
super-adiabatic thermal boundary layer is assumed that inhibits but not totally suppresses the particle
and heat exchange. We find that the increment in temperature across the TBL must not exceed 1300--4000 K
at a pressure level of 10 GPa in order to keep the deep adiabat cold enough. The change in temperature
could perhaps be as small as a few 100 K if the TBL delayed the particle transport and thus the depletion
of the atmosphere. Without that effect values of 1300--4000~K are favored depending on 
the H/He phase diagram, EOS, and deep water abundance. 
A TBL of the order of a few 100 to a few 1000 K is consistent with the high luminosity of
Neptune if it established a few 100 Myr after its formation \citep{Scheibe21}.
In this case Ne/H could be depleted as well, depending on the onset of He-rain and the timescale 
of neon diffusion compared to helium diffusion through the TBL.

Even though the 1-bar temperatures of Uranus ($72\pm 2$ K) and Neptune ($76\pm 2$ K) are very similar
as compared to those of Saturn (135 K) and Jupiter (170 K), our models with a TBL that yield weak
He depletion for Uranus predict a stronger depletion for Neptune due to this small 4 K difference.
This is in contrast to observations, which show a higher He abundance in the atmosphere of Neptune
\citep{Atreya20}. The different atmospheric He abundances may be a sign of different internal
structures \citep{HF20}.

\item
\textit{Atmospheric He enrichment.}
An atmospheric  enrichment in helium might be possible if phase separation processes take place that
affect hydrogen and helium differently.
Phase separation of water from hydrogen has been suggested to explain the reduced heavy element
abundances in the outer envelopes of Uranus and Neptunes as revealed by their observed gravity
data \citep{Bailey21}. Rain-out of water would reduce the amount of water in the deep atmosphere
down to a level which is dictated by the H$_2$/H$_2$O phase diagram but leave the atmospheric He/H ratio
unaffected.  Simultaneously, H$_2$ would rise from the deep interior upward until the concentration
of the remaining H$_2$ is low enough to be miscible in the ionic water environment \citep{Bailey21}.
Limited solubility of H$_2$ gas in salty water is also seen in laboratory experiments that were conducted
to investigate the storage of gaseous H$_2$ in underground water reservoirs \citep{Chabab20},
which happens to be at $P$--$T$ conditions of $\sim 200$ bars and $\sim 350$ K along the Neptune adiabat.
Rising H$_2$ would enrich the atmosphere and let it appear depleted in helium. However, the miscibility
of helium in ionic water is not so well studied, and helium may rise as well.

An enrichment in He could also be seen if large amounts of CO or other carbon-bearing species were absorbed
by the planet. These could then react with H$_2$ in the hydrogen-environment to form methane and water.
Whether the amount of H$_2$ particles that can be bound in form of volatiles is sufficient to deplete
the atmosphere in H$_2$ and let it appear He-enriched remains to be estimated.

An enrichment in He could also be possible if He atoms prefer to partition into the molecular-hydrogen
layer before forming droplets that would rain-out because of their higher mean molecular weight 
\citep{SS77a,SS77b}. To address the behavior of He once the Gibbs free energy difference suggests phase 
separation may be studied in direct computer simulations with He atoms in molecular and metallic hydrogen, 
respectively.

One may think that a He enrichment could occur due to late formation in an evaporating protosolar
disk, as the lightest elements may escape first and thus leave behind a gas component that is 
primarily depleted in hydrogen. In that scenario however, the apparent enrichment in the noble gases should
increase with their molecular weight. \citet{GuiHue06} showed that the escape of disk gas would occur
hydrodynamically without separation of the species in the disk gas. The reason why the remaining
disk would be equally enriched in the heavy noble gases Ar, Kr, Xe, and perhaps explain their
 enrichment in the atmosphere of Jupiter, is that these elements can be captured by grains that 
settle onto the disk midplane, where the planets form.
\end{itemize}

\subsection{Gravity}

The uncertainty in interior structure and composition is partially due to the magnitude in the 
uncertainty of the observational constraints and their limited number. For instance, current interior 
models of the outer planets are constrained by the low-order gravitational harmonics $J_2$ and $J_4$. 
The uncertainty in the $J_4$ values of Uranus and especially of Neptune allows for a wide 
of interior models \citep{Kaspi13,Nettelmann13,Movshovitz22}.
Knowing the depth of the zonal winds would constrain the dynamic correction to $J_4$ and thus could eliminate
a large fraction of the models \citep{Kaspi13}. Deep winds at depths of the order of 1000 km also change 
the higher order moments $J_6$ and $J_8$ by a significant amount of 20\% and 30\%, respectively 
\citep{Neuenschwander22}. \citet{Movshovitz22} also showed that $J_6$ 
and $J_8$ constrain the range of deep interior density distribution much tighter for the ice giants 
than they do for Jupiter and Saturn. It is to be expected that a measurement of the higher-order moments 
$J_6$ and $J_8$ for Uranus and Neptune would reduce the current variety of possible models, 
both directly and through inference of the wind depth. Such measurement can only be achieved by an Orbiter mission.

\section{Summary}\label{sec:summary}

We have applied three H/He phase diagrams (LHR0911,SR18,BLM21-N23) and three H/He-EOS (CMS19,CD21,REOS.3)
and assumed a wide range of deep water abundances (0--0.9) as well as the presence of a thermal boundary layer
(TBL) for the ice giants to address He depletion in the outer giant planets. Our conclusions are as follows.

\begin{enumerate}
\item
We expect to see strong depletion in He in the atmospheres of both Uranus and Neptune
if their deep interiors are adiabatic and contain H/He.
\item
The observed weak helium depletion in the atmosphere of Uranus might be indicative of a moderate
TBL and the presence of H/He in its deep interior.
\item
The observed protosolar He/H ratio in the atmosphere of Neptune might be indicative of a helium-free
deep interior composed of ices and rocks.
\item
If the interiors of Uranus and Neptune are similar, our models predict a slightly stronger
helium depletion for Neptune than for Uranus, contrary to what is observed. This suggests their 
interiors are dissimilar to some extent.
\item
Considering atmospheric enrichment in helium is beyond the scope of this work but could happen 
in the Ice Giants as a result of hydrogenization of abundant CO.
\item
For Saturn, we expect any He depletion as a result of uncertainties in the interior structure,
although most attempts to reconcile a potentially measured value with the jovian value given a H/He
phase diagram requires the presence of an outer stable layer (OSL) in both planets.
\item
For Jupiter, all considered H/He phase diagrams may explain its observed H/He depletion but all
require the presence of an outer OSL. Cold demixing curves (SR18) require a sub-adiabatic OSL.
\item
Explaining both Jupiter and Saturn with the same H/He phase diagram seems problematic if their
interiors are adiabatic. Conversely, this issue is immediately lifted if an OSL is allowed for.
\item
More experimental constraints on the H/He phase diagram are urgently needed given the powerful
implications of the phase diagram on the interior structure of the giant planets, especially
in conjuction with a measured atmospheric He abundance.
\item
While here we focused on helium only, an in-situ measurement of the abundances of Ne and heavier
noble gases in addition to that of He would tremendously help to inform us about the
deep composition and the atmosphere-interior coupling in the giant planets. 
\end{enumerate}
The science addressed in this work requires improved knowledge of the H/He phase diagram
and an entry probe down to the levels of $\sim$5 bars that is equipped with a mass spectrometer to
one of the ice giants but best also to Saturn. The science return from an entry 
probe does not depend critically on a launch window. Hence an Uranus Entry Probe-centered mission
design such as the UOP \citep{Mandt23} may compete against lighter mission designs that trade 
the science objectives for lower cost and earlier launch time \citep{Cohen22}.

\backmatter

\bmhead{Acknowledgments}

N.N.~acknowledges support through NASA's Juno Participating Scientist Program under grant
80NSSC19K1286. This work was also supported by the DFG Research Unit FOR2440/2 under
grant NE1734/2-2. We thank Armin Bergermann for providing the tabulated data of the SR18
phase diagram and the two referees for constructive comments that helped us to improve 
the paper.

\section*{Declarations}

\begin{itemize}
\item Funding:
Partial financial support was received from NASA under Grant 80NSSC19K1286 and from the German Science Foundation (DFG) under Grant NE1734/2-2. 

\item Conflict of interest: 
The authors have no competing interests to declare that are relevant to the content of this article.

\end{itemize}


\begin{appendices}

\section{Construction of the BLM21-N23 phase diagram}\label{sec:apxBLM}

This Section reports our construction of a H/He phase diagram that is based on the experimental data 
of \citet{Brygoo21}. We label it BLM21-N23 and have constructed it as simple as described below.

\citet{Brygoo21} provide a demixing curve for 11\% He mol fraction. The curve interpolates between four important
experimental data points in $P$--$T$ space. At the point at $P=0.93$ Mbar and $T=4700 \pm 200$K 
the reflectivity is found to increase with increasing pressure to levels that are found to be
characteristic of a H-rich/He-poor phase, while at the point at ($P=1.5$ Mbar,$T=10.400\pm 500$ K) the
reflectivity decreases. These two points are interpreted to mark, respectively, the 
entry into and leave from  the demixing region. We have estimated the here given uncertainty 
in $T$ from the figures in \citep{Brygoo21}. Toward higher pressures, they extend the demixing curve  
 as a flat line up to $\sim$2.5 Mbar. Toward lower pressures threy draw the demixing curve to run
smoothly through two further experimental data points of \citep{Loubeyre87} at 6 and 9 GPa, respectively.

First, we construct a demixing line for $\xHe=0.05$ mol fraction between 0.6 and 2 Mbar. Second, 
we use that curve to construct further demixing curves $\Tdmx(\xHe;P)$ by linear interpolation in 
$T(\xHe)$ at a number of pressure points. Third, we assume a symmetric shape of the phase diagram 
$\Tdmx(\xHe;P)$ with respect to the dividing line at $\xHe=0.5$. For all pressures, 
we set $\Tdmx(\xHe=0.5) = \Tdmx(\xHe=0.11)$. This ensures a monotonic 
behavior in $\Tdmx(\xHe)$ in lack of experimental data. For comparison, the increase
in the SR18 data along the 2 Mbar isobar over this concentration range is 3300 K while for the LHR0911 data
it is 1300 K. However, the $\xHe=0.11$ mole fraction curve corresponds to $Y=0.33$, which is 
more He-rich than in Jupiter's deep interior according to models with a He-poor outer envelope of the 
Galileo entry probe He abundance of $Y=0.238 \pm 0.005$. Jupiter models do not reach into this 
high-He abundance region. On the other hand in Saturn's deep interior the He-abundance can easily 
reach levels of $Y\sim0.4$ or higher \citep{Nettelmann13,MF20} and thus exceed the region of 
our version of the BLM21 phase diagram where it is supported by data.

\begin{center}
\begin{figure}
\rotatebox{270}{\includegraphics[width=0.68\textwidth]{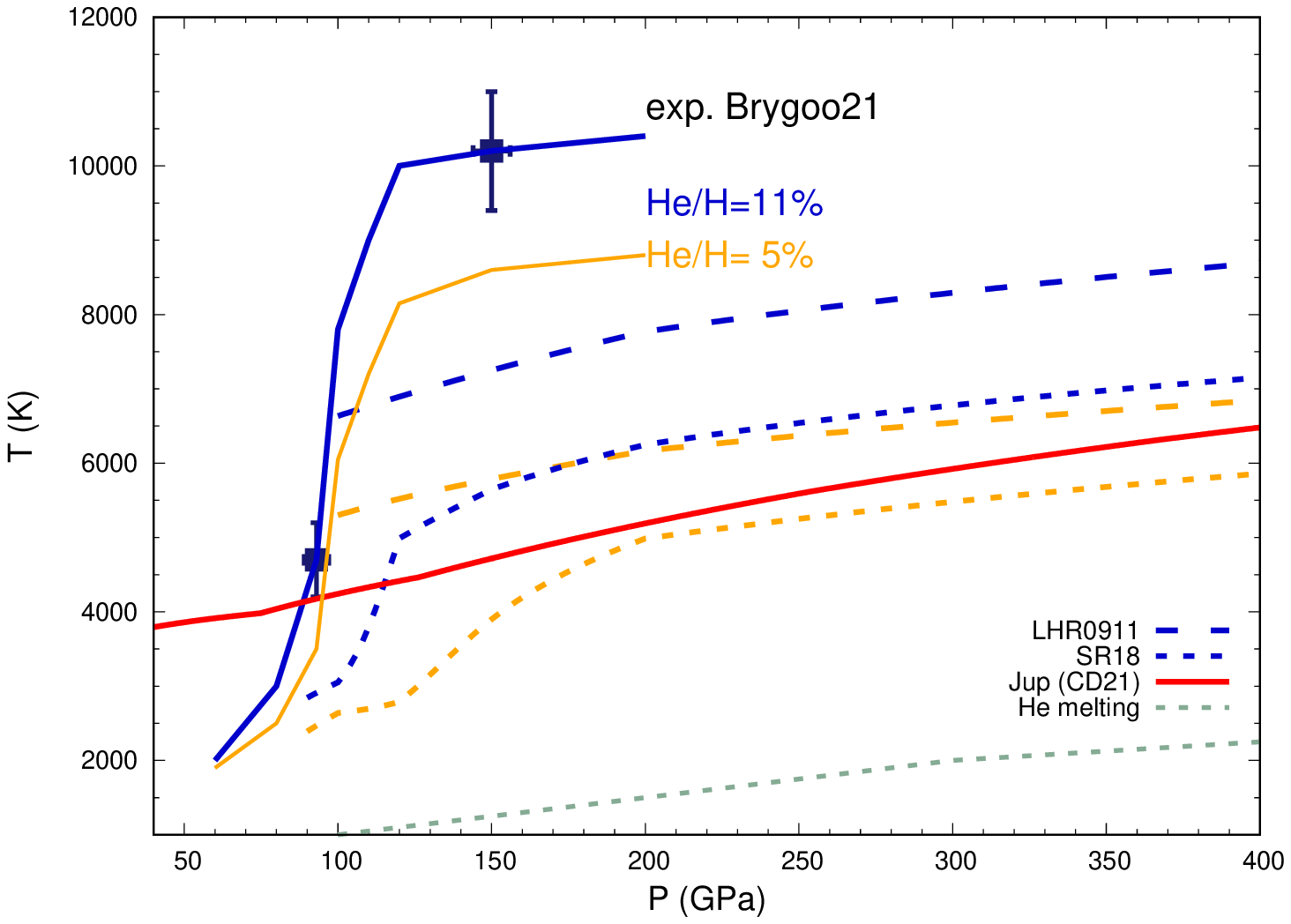}}
\caption{\label{fig:BLM21}
Constructed H/He-demixing curve at $\xHe=0.05$ (solid orange line) to obtain the BLM21-N23 H/He 
phase diagram based on the BLM21-demixing line at $\xHe=0.11$ (solid blue line) and the 
approximate temperature differences at 1.5--2 Mbars in the LHR0911 (dashed) and the SR18 (dotted) data  
for these two He abundances. The grey-green curve is the He melting line of Ref.~\citep{Preising19} while 
the red curve is a Jupiter model based on CD21-EOS (Nettelmann et al in prep).  
}
\end{figure}
\end{center}

To obtain the curve for $\xHe=0.05$ we use the SR18 and LHR0911 data as guidance. At 1.5 Mbars, 
the temperature difference between $\xHe=0.11$ and 0.05 is $-1750$K (SR18) and $-1500$K (LHR0911), 
while at 2 Mbars the difference is $-1260\:$K and $-1600\:$K, respectively. We assume an intermediate difference 
of $-1600$ K at 1.5 Mbar, which because of the flat behavior in the $\xHe$=0.11 curve of \citet{Brygoo21} 
is also assumed for 2 Mbars and beyond. Toward lower pressures we let the temperature difference decrease to 
$-100\:$K at our lowermost pressure of 0.6 Mbar. For comparison, the difference at 0.5 Mbar in the SR18 
data is $-350\:$K. Our choice of $-100\:$K ensures that the $\xHe=0.05$ curve would not intersect 
with the SR18 curve upon extrapolation but approach it toward lower pressures. 
At intermediate pressures, we let the temperature decrease by amounts that are chosen 
to yield a well-behaved curve, which is displayed as the orange line in Figure \ref{fig:BLM21}.

On the low-$P$ end, all demixing lines are well above the He melting line  \citep{Preising19}. We also
recover the behavior that Jupiter adiabats, if they are cold enough depending on the H/He-EOS used, 
will intersect with the SR18 protosolar abundance demixing line near 2 Mbars while for the LHR0911 
data, at 1 Mbar. Interestingly, the intersection of the displayed Jupiter adiabat, here computed 
using the CD21-EOS, with the BLM21-N23 diagram can also be expected to occur at $\sim$0.9 Mbar.
At 0-9--1 Mbars, the demixing temperatures are not that much higher than for the LHR0911 data and 
therefore, we expect a not that much lower inferred level of He depletion, consistent with our
results in Figure \ref{fig:YatmT_LHR}.

However we caution that our construction of a BLM-N23 H/H phase diagram is not based on the 
principle of thermodynamic consistency of an underlying H/He EOS but on two reflectivity data points
with high experimental uncertainty. More reflectivty measurements and theoretical predictions for
H/He mixtures are needed to evaluate such an approach.

\end{appendices}

\bibliography{refs_IceGiants}

\end{document}